\documentstyle[aps]{revtex}
\begin{document}
\newcommand{\beq}{\begin{equation}}
\newcommand{\eeq}{\end{equation}}
\newcommand{\beqn}{\begin{eqnarray}}
\newcommand{\eeqn}{\end{eqnarray}}
\newcommand{\bmath}{\begin{mathletters}}
\newcommand{\emath}{\end{mathletters}}
\twocolumn[\hsize\textwidth\columnwidth\hsize\csname @twocolumnfalse\endcsname
\title{Metallic ferromagnetism without exchange splitting}
\author{J. E. Hirsch }
\address{Department of Physics, University of California, San Diego\\
La Jolla, CA 92093-0319}
 
\date{\today} 
\maketitle

\begin{abstract} 

In the band theory of ferromagnetism there is a relative shift in the position of majority
and minority spin bands due to the self-consistent field due to opposite spin electrons. 
In the simplest realization, the Stoner model, the majority and minority
spin bands are rigidly
shifted with respect to each other.  Here we consider models at the opposite extreme, where
there is no overall shift of the energy bands.  Instead, upon spin polarization one of the
bands broadens relative to the other. Ferromagnetism is driven by the
resulting gain in kinetic energy. A signature of this class of
mechanisms is that a transfer of spectral weight in optical absorption
from high to low frequencies occurs upon spin polarization. We show that such
models arise from generalized tight binding models that include off-diagonal matrix
elements of the Coulomb interaction. For certain parameter ranges it
is also found that reentrant ferromagnetism occurs.
 We examine properties of these models at zero and finite temperatures, and
discuss their possible relevance to real materials. 

\end{abstract}
\vskip2pc]

\section{Introduction}
In the Stoner model for metallic ferromagnetism\cite{ston}, a rigid shift in the
positions of the majority and minority spin bands occurs.  The difference in
band energy between minority and majority spin electrons is:
\beq
\epsilon _{k\downarrow} - \epsilon_{k\uparrow} = \Delta = U m
\eeq
Here $\Delta$ is the exchange splitting, $m$ the magnetization and $U$ the exchange
interaction. In deriving the Stoner model from a tight binding model, $U$
corresponds to an on-site interaction,because the shift in Eq.(1) is momentum-
independent. The Stoner model has been widely used for the description of
metallic ferromagnets\cite{wolf}.

More elaborate approaches using spin density functional theory derive the
band structure of majority and minority spins taking into account
electron-electron interactions within the local spin-density 
approximation\cite{moru}.
The result is exchange-split bands where the exchange splitting is in general not
constant across the Brillouin Zone.  Nevertheless, here again there is always an
overall shift in energy of majority and minority spin bands with respect to
each other.

In this paper we consider models where there is no overall shift in the relative
position of the energy bands.  Can ferromagnetism still occur?  The answer is yes, if
there is instead a change in the relative bandwidth of majority and minority spin
electrons upon spin polarization.  Figure 1 contrasts the situation
 in the Stoner model and in the models considered in this paper. The gain
of energy in the ferromagnetic state arises from an effective mass reduction,
or equivalently a bandwidth expansion, of the majority spin electrons, leading
to a lowering of kinetic energy.

The possible  role of effective mass reduction in
driving the transition to ferromagnetism was first proposed in
connection with the 'double exchange' mechanism to describe
manganese oxides\cite{doub}. More generally, we have suggested in the
context of describing metallic ferromagnetism with a single band
generalized tight binding model\cite{hirs1,hirs2}, that effective mass
reduction upon spin polarization may be an essential feature
of $all$ itinerant ferromagnets. It was pointed out\cite{hirs1} that
the anomalous drop in resistivity and the negative magnetoresistance
that is usually found in ferromagnets could be explained as arising from a change in
effective mass rather than a change in scattering time as usually
assumed\cite{dege}. Clear experimental evidence for this
phenomenon has been found in optical properties of the colossal
magnetoresistance manganites\cite{okim} as well as of some rare-earth
hexaborides\cite{degi} and magnetic semiconductors\cite{dimi}. 
Whether this is a universal phenomenon
that occurs in all metallic ferromagnets remains an open question.

In the model that we considered originally\cite{hirs1}, ferromagnetism arises
from the combined effect of exchange energy and band broadening as
the system orders. Such may well be the situation applicable to
real materials. However, to better understand the various theoretical
possibilities it is of interest to ask whether ferromagnetism could be 
driven just by band broadening in the absence of exchange energy. As we show in 
this paper, this is indeed possible, if the bandwidth for one spin
expands while the one for the other spin contracts. In contrast, in the
model considered originally where exchange energy was also
present, both minority and majority spin bands could expand
upon spin polarization. Still, both in the original model and in
the ones considered here a signature of this physics is that an overall
shift in optical spectral weight from high to low frequencies takes place as
spin polarization develops. Application of this model to describe
the shift in optical spectral weight observed in $EuB_6$ is discussed
elsewhere\cite{hirs5}.

\section{The Models}

We consider a single band of width D, with density of states $g(\epsilon)$
symmetric around the origin:
\beq
\int_{-D/2}^{D/2} d\epsilon g (\epsilon ) \epsilon =0
\eeq
The magnetization $m$ and number of electrons $n$ per site are given by
\bmath
\beq
m = \int_{-D/2}^{D/2} d{\epsilon } g (\epsilon ) [f (\epsilon _\uparrow(\epsilon )) -
f(\epsilon _\downarrow(\epsilon ))]
\eeq
\beq
n = \int_{-D/2}^{D/2} d{\epsilon } g (\epsilon ) [f (\epsilon _\uparrow(\epsilon )) +
f(\epsilon_\downarrow(\epsilon ))]
\eeq
\emath
with $f$ the Fermi function.  In the class of models under consideration here
the  quasiparticle energies are given by
\beq
\epsilon _\sigma(\epsilon ) = (1-2j_1I_\sigma - 2j_2 I_{-\sigma})\epsilon 
-\sigma D [\frac{k}{2} m + h] -\mu
\eeq
Here, $h=H/D$, with H an external magnetic field (in units of energy), and the bond
charge $I_\sigma$ is given by
\beq
I_\sigma = \int_{-D/2}^{D/2} d\epsilon g (\epsilon ) (\frac{-\epsilon }{D/2})
f(\epsilon _\sigma(\epsilon ))
\eeq
The ``exchange'' interaction $k$ gives an overall rigid shift in the band energies, as in
the Stoner model.  The new features here arise from the interactions $j_1$ and $j_2$ that give
rise to band narrowing that depends on the bond-charge occupation for each spin $I_\sigma$,
which in turn will be a function of temperature and magnetization.

\section{Relation to tight binding models}

Consider a tight binding model with on-site, nearest-neighbor, exchange and pair hopping
interactions\cite{hubb,kive,camp,hirs2}:
\beqn
H = &-& t \sum _{<ij>,\sigma}(c_{i\uparrow}^\dagger c_{j\sigma} + h.c.) +
U\sum_{i}n_{i\uparrow} n_{i\downarrow}\nonumber \\
&+&V\sum_{<ij>}n_in_j +J\sum_{<ij>,\sigma,\sigma'}
c_{i\sigma}^\dagger c_{j\sigma'}^\dagger c_{i\sigma'}c_{j\sigma}\nonumber \\
&+&J'\sum_{<ij>}(c_{i\uparrow}^\dagger c_{i\downarrow}^\dagger
c_{j\downarrow}c_{j\uparrow}+h.c.)
\eeqn
The interactions arise from the various matrix elements of the Coulomb
interaction between orbitals at nearest neighbor sites.  A mean field
decoupling leads to the quasiparticle energies
\beq
\epsilon _\sigma(k)=[1-I_\sigma(\frac{J}{t}-\frac{V}{t})-
I_{-\sigma}(\frac{J}{t}+\frac{J'}{t})]\epsilon _k-
\sigma\frac{(U+zJ)}{2}m
\eeq
with
\bmath
\beq
I_\sigma=<c_{i\sigma}^\dagger c_{j\sigma}>
\eeq
\beq
m=<c_{i\uparrow}^\dagger c_{i\uparrow}>-<c_{i\downarrow}^\dagger c_{i\downarrow}>
\eeq
\emath
the bond charge and magnetization parameters.  The bandwidth and band energies are
\bmath
\beq
D=2zt
\eeq
\beq
\epsilon_k=-t\sum_\delta e^{ik\delta}
\eeq
\emath
with z the number of nearest neighbors to a site and $\delta$ a vector connecting
a site to its nearest neighbors. The interactions in Eq. (4) are then
\bmath
\beq
j_1=\frac{zJ}{D}-\frac{zV}{D}
\eeq
\beq
j_2=\frac{zJ}{D}+\frac{zJ'}{D}
\eeq
\beq
k=\frac{U}{D}+\frac{zJ}{D}
\eeq
\emath
In connection with itinerant ferromagnetism we have considered in the past the model with 
$U$ and $J$ only\cite{hirs1}, and also the one with $U$, $J$  and $J'=J$\cite{hirs3}.  
In terms
of the notation in Eq. (4), the first case corresponds to
\bmath
\beq
j_1=j_2=j
\eeq
\beq
k=u+j
\eeq
\emath
with $j=zJ/D$, $u=U/D$. In this particularly simple case the 
hopping amplitude gets renormalized equally by bond charge density
of equal and opposite spin. When one includes the pair hopping term 
$J'$, and in particular for $J'=J$, we have
\bmath
\beq 
j_1 = j 
\eeq

\beq
j_2 = 2j
\eeq

\beq
k = u+j
\eeq
\emath
so that in this case more of the hopping renormalization is due to electrons of opposite
spin.  As seen in Ref. 15, this case favors more the existence of ferromagnetism
near the top and bottom of the band compared to the first case.

Note that according to Eq. (10) the band renormalization parameter $j_1$ from the same-spin
electrons should be expected to be smaller than that of opposite spin electrons, $j_2$, due to
the effects of nearest neighbor repulsion and of pair hopping.  In this paper we
will focus on that situation.

\section{Ground State Properties}
We parametrize the interactions in Eq. (4) as

\bmath
\beq
j_1 = aj
\eeq

\beq
j_2 = j
\eeq
\emath
with $0\leq a \leq 1$.  The case studied previously with only exchange
interaction $J$ corresponds to $a=1$, and the case with pair hopping $J'=J$
corresponds to $a=1/2$.  If the system is not completely polarized the chemical
potential is determined by the condition
\beq
\epsilon _\uparrow(\epsilon _{F\uparrow})=
\epsilon _\downarrow(\epsilon _{F\downarrow})
\eeq
with
\beq
n_\sigma=\int_{-D/2}^{\epsilon _{F\sigma}}d\epsilon g(\epsilon)
\eeq
and
\bmath
\beq
n=n_\uparrow+n_\downarrow
\eeq
\beq
m=n_\uparrow-n_\downarrow
\eeq
\emath
the total occupation and magnetization respectively.  We assume for simplicity a flat
density of states
\beq
g(\epsilon )=\frac{1}{D}, -\frac{D}{2}\leq \epsilon \leq \frac{D}{2}
\eeq
We have seen earlier that the properties of these type of models do not depend strongly on
energy variation of the density of states, unlike in the Stoner model.

The occupations of up and down spin bands at $T=0$ are given by
\beq
n_\sigma=\frac{\epsilon _{F\sigma}}{D}+\frac{1}{2}
\eeq
so that
\beq
\frac{\epsilon _{F\sigma}}{D}=\frac{n+\sigma m-1}{2}
\eeq
and the bond charge densities are given by
\beq
I_\sigma=\frac{1-(1-n-\sigma m)^2}{4}=n_\sigma(1-n_\sigma)
\eeq
so that the condition Eq. (14) is
\beq
(j_1-j_2)(1-n)^2-(j_1 +j_2)\frac{1-m^2-(1-n)^2}{2}=k-1
\eeq
and for the parametrization Eq. (13) we obtain
\beq
j=\frac{2(1-k)}{(1+a)(1-m^2)-(3a-1)(1-n)^2}
\eeq
The conditions for onset of ferromagnetism and for full spin polarization result from
setting  $m=0$ and $m=n$ respectively in Eq. (22).

The resulting phase diagrams for various values of $a$ are shown in Fig. 2.  For all cases,
partially polarized ferromagnetic regions occur predominant around the 1/2-filled
band.  Note that for exactly half-filling no full polarization ocurs for $k<1$. Also,
note that for small values of $a$, spin polarization is easier near the edges of the band,
and for large values of $a$ near the half-filled band. For the value $a=1/3$ (not shown),
the boundary for onset of spin polarization is independent of $n$, at
$j/(1-k)=1.5$.

However, the phase diagrams in Fig. 2 need to be supplemented by the condition that the
coefficients of $\epsilon$ in Eq. (4) need to remain positive, i.e.
\beq
1-2j_1I_\sigma-2j_2I_{-\sigma}>0
\eeq
which may not be satisfied for some values of $k$. While this condition is always
satisfied at the points where onset of ferromagnetism occurs, it is not so
for partial or full spin polarization.
For $k=0$, condition (23) is equivalent to the constraint
\beq
n\leq 1-m
\eeq
for $n\leq1$, and the corresponding symmetric one for $n>1$.  In particular, condition (24)
implies that full polarization in the absence of exchange, i.e. $k = 0$, can only be achieved
for $n \leq 1/2$ or $n\geq 3/2$.  Partial polarization of magnitude $m = cn$, $0<c<1$,
can be achieved only in the range $n<1/(1+c)$ and the corresponding
symmetric one for $n>1$.

For non-zero exchange $k$, full polarization can be achieved for any n if the
condition
\beq
\frac{k}{1-a}\geq \frac{1}{2}
\eeq
holds, which in particular is true if $k\geq 0.5$ for any $a$. 
(By contrast, the Stoner model requires $k>1$ to give rise to
ferromagnetism.) Otherwise, full
polarization is only achieved in the range
\beq
n\leq \frac{1/2}{1-\frac{k}{1-a}}
\eeq
and the corresponding symmetric one above half-filling.  Note that as $a$ approaches unity,
condition (26) will be satisfied for any filling for arbitrarily small exchange $k$.

Note also that the case $a = 1$ is a very singular point.  For exchange $k=0$, there is no
ferromagnetism in the model as no change in relative occupation of up and down spin
bands can occur.  On the other hand, for any $k\neq 0$ the criteria for onset and full
polarization can be satisfied for suitable $j$ for any value of occupation $n$.

\section{Finite temperatures}

Eq. (5) for the bond charge yields
\beq
I_\uparrow-I_\downarrow = 
\int_{-D/2}^{D/2} d{\epsilon } g (\epsilon ) (\frac{-\epsilon }{D/2})
[f(\epsilon _\uparrow(\epsilon )) - f(\epsilon _\downarrow(\epsilon ))]
\eeq
and using
\beq
j_1I_\uparrow+j_2I_\downarrow=\frac{j_1+j_2}{2}(I_\uparrow+I_\downarrow)
+\frac{j_1-j_2}{2}(I_\uparrow-I_\downarrow)
\eeq
we obtain on expanding the Fermi functions on the right-hand side
of Eq. (27):
\beq
I_\uparrow-I_\downarrow = (I_\uparrow-I_\downarrow)(j_2-j_1)D
 \int_{-D/2}^{D/2} d\epsilon g
(\epsilon ) (\frac{-\epsilon }{D/2})^2 (-\frac{\partial f}{\partial \epsilon _\sigma})
\eeq
in the absence of exchange, $k=0$. Thus the equation that determines the
critical temperature is found by cancelling $(I_\uparrow-I_\downarrow)$ on
both sides of this equation:
\beq
1 = (j_2-j_1)D
 \int_{-D/2}^{D/2} d\epsilon g
(\epsilon) (\frac{-\epsilon }{D/2})^2 (-\frac{\partial f}{\partial \epsilon _\sigma})
\eeq
More generally, in the presence of exchange we consider also Eq. (3a) for the
magnetization, and obtain the set of equations
\bmath
\beq
I_\uparrow-I_\downarrow=(j_2-j_1)(I_\uparrow-I_\downarrow)G_2+kmG_1
\eeq
\beq
m=(j_2-j_1)(I_\uparrow-I_\downarrow)G_1+kmG_0
\eeq
\emath
with
\beq
G_l=D\int_{-D/2}^{D/2} d\epsilon g
(\epsilon ) (\frac{-\epsilon }{D/2})^l (-\frac{\partial f}{\partial \epsilon _\sigma})
\eeq
and the condition for the critical temperature is obtained by
setting to zero the determinant of the coupled Eqs. (31), yielding
\beq
[1-(j_2-j_1)G_2][1-kG_0]-(j_2-j_1]kG_1^2=0
\eeq

We focus here on ferromagnetism without exchange, so that the
$T_c$ equation is given by Eq. (30), or
\beq
1=(j_2-j_1)G_2   .
\eeq
For infinite temperature $G_2$ is zero, and as $T$ decreases it
becomes positive. At a critical temperature, that increases when the
difference $(j_2-j_1)$ increases, Eq. (34) will be satisfied.
The situation is analogous to the usual Stoner model where the
critical temperature is determined by the equation 
\beq
1=kG_0    .
\eeq
At low temperatures we can approximate the Fermi function derivative 
in Eq. (34) as a $\delta$-function, yielding \beq
1=\frac{4(j_2-j_1)}{1-2(j_1+j_2)I_\sigma}
(\frac{\epsilon _{F\sigma}}{D})^2
\eeq
with $(\epsilon _{F\sigma}/D)$ given by Eq. (19). As
$T\rightarrow0$, $I_\sigma$ is given by 
Eq. (20) and the $T_c$ equation (36) reduces to the condition
Eq. (22) for onset of ferromagnetism ($m=0$) for the case $k=0$.
One can also obtain an approximate analytic form for $T_c$ in 
weak coupling by expanding $G_2$ in Eq. (34) to one higher
order than used to obtain Eq. (36). This is in contrast to the
case $j_1=j_2$, where the lowest order equation analogous to
Eq. (36),
\beq
1=\frac{k}{1-2j(I_\uparrow+I_\downarrow)}
\eeq directly yields $T_c$ at low temperatures on expanding
$I_\sigma$ as
\beq
I_\sigma=I_\sigma^0-\frac{\pi^2}{3}(\frac{k_BT}{D})^2
\frac{1}{1-2(j_1+j_2)I_\sigma^0}
\eeq
with $I_\sigma^0=I_\sigma(T=0)$.

In fact, the integral $G_2$ (Eq. (32)) will not 
 be monotonically decreasing as the temperature
increases, as $G_0$ is. For example, for a half-filled
band the Fermi function derivative as $T\rightarrow 0$
approaches a $\delta$-function at zero energy, and the
integral $G_2$ will vanish due to the extra factors
of $\epsilon$. As the temperature is increased
from zero, $G_2$ will increase until a critical 
temperature is reached where Eq. (34) is satisfied.
At a higher temperature $G_2$ will start decreasing
until a second temperature is reached where Eq. (34) is satisfied.
Thus we can expect to find re-entrant ferromagnetism in this
model for intermediate values of the effective coupling $(j_2-j_1)$.

The magnetic susceptibility above $T_c$ is obtained by taking
the derivative of the magnetization Eq. (3a) with respect to
the external magnetic field. For the case $j_1\neq j_2$ the magnetic
field dependence of the bond charge needs to be taken into account.
 We obtain for
$\chi=dm/dh$ \beq
\chi=2\frac{G_0-(j_2-j_1)(G_0g_2-G_1^2)}
{(1-kG_0)[1-(j_2-j_1)G_2)]-k(j_2-j_1)G_1^2}
\eeq
with $G_l$ given by Eq. (32). In particular, $G_0$ is
the
magnetic susceptibility per spin in the absence of
interactions, and Eq. (39) reduces to the usual RPA form 
if $j_1=j_2$.\cite{hirs1} More generally, the susceptibility
Eq. (39) diverges as $1/(T-T_c)$ as $T$ approaches
$T_c$ given by the solution of Eq. (33). 
Parametrization of the susceptibility as
\beq
\chi(T)=\frac{p_{eff}^2(T)}{3(T-T_c)}
\eeq
defines the effective moment $p_{eff}$, which is 
temperature-independent within the Curie-Weiss
law, which is often seen experimentally.

\section{Numerical results}

We solve the mean field equations numerically for various cases.
Figure 3 shows the magnetization versus temperature for $n=0.5$, $a=0$
(so that $j_1=0$) and various values of $j_2=j$ in the absence of
exchange ($k=0$). In this
case the phase diagram Fig. 2a shows that onset of magnetization
at zero temperature occurs for $j=1.6$ and full polarization
for $j=2$. However note that at finite temperature ferromagnetism
also occurs for $j<1.6$ as discussed in the previous section.
For $j=1.6$ the magnetization approaches zero as
$T\rightarrow 0$, and for $j<1.6$ reentrant ferromagnetism occurs.
For $j=2$ the system becomes fully polarized as $T\rightarrow 0$,
and for $j>2$ the model becomes unphysical because the condition
Eq. (23) fails to be satisfied as the temperature is lowered.
Note that for these cases the shape of the magnetization curve is
very different than what is obtained from the Stoner model or 
from molecular field theory.

The effective bandwidth for spin $\sigma$, $D_\sigma^*$, or
equivalently the effective mass $m_\sigma^*$, are given by the
relation
\beq
\frac{D}{D_{\sigma}^*}=\frac{m_{\sigma}^*}{m}
=1-2j_1I_{\sigma}-2j_2I_{-\sigma}
\eeq
with $D$ and $m$ the (spin-independent) bare bandwidth and mass
respectively. Figure 4 shows the temperature dependence.
For the majority spins, the effective mass increases as the
temperature decreases, and then decreases as $T$ is lowered below
$T_c$. For the minority spins, $m^*$ increases more sharply as
$T$ is lowered below $T_c$. 

For small values of $j$ however the effective mass for the 
majority spins also increases as the temperature is lowered 
further, and becomes eventually larger than its value
at $T_c$. One may ask why it is still advantageous to
spin-polarize, when the energy will no longer be lower than
what it was in the unpolarized state at $T_c$. The answer is 
of course that at this lower temperature if the system was
unpolarized the bandwidth for both up and down spin
electrons would be much smaller, giving rise to a larger
energy than in the polarized state (see Fig. 1 of
Ref. 5(b) for an illustration of this effect).

The behavior of the effective moment, Eq. (40), versus temperature
for this case is shown in Fig. 5. For small $j$ the temperature dependence
is strong and it increases as $T$ approaches $T_c$, for larger $j$
it decreases somewhat as $T$ approaches $T_c$.

Results for a case where the band renormalization for the same
spin, $j_1$, is not zero, are shown in Fig. 6. The case 
$a=0.5$ shown corresponds to parameters $J'=J$ and $V=0$ in the
tight binding model. Here the effective moment is much less
temperature dependent and no reentrant behavior is found for
this set of parameters.

Fig. 7 shows the effect of exchange on the behavior of 
the magnetization and effective moment. With increasing $k$ the
magnetization curve becomes steeper and resembles more the
conventional behavior. The effective moment is rather
constant with temperature for these cases corresponding
to a Curie-Weiss law for the susceptibility. As discussed
in Ref. 5, for the pure Stoner model (exchange only) a strong
temperature dependence of the effective moment is found.

\section{Optical absorption, magnetoresistance and photoemission}

The Drude formula for the real part of the optical conductivity is
\beq
\sigma_1(\omega) = \frac{ne^2}{m^*} \frac {\tau}{1+\omega^2\tau^2}
\eeq
which describes optical absorption through intra-band processes. The conductivity
sum rule that results from this formula is
\beq
\int_0^{\omega_m} d\omega \sigma_1(\omega)=\frac{\pi}{2}\frac{ne^2}{m^*}
\eeq
The cutoff $\omega_m$ is to exclude transitions to other
bands, which are processes not described by the expression
Eq. (42). We assume that $\omega_m \tau>>1$ in order to derive
Eq. (43) from Eq. (42). Ordinarily, if no change in effective
mass occurs upon spin polarization, the optical absorption
given by these formulas will not depend on spin polarization
(assuming a constant $n$). While one may expect changes in 
optical absorption as function of temperature due to changes
in the relaxation time $\tau$, no changes would be expected as
function of magnetic field at fixed temperature.

In contrast, in the models considered here we have
\beq
\sigma_1(\omega)=\frac{e^2\tau}{1+\omega^2\tau^2}
(\frac{n_\uparrow}{m^*_\uparrow}+
\frac{n_\downarrow}{m^*_\downarrow})
\eeq
and
\beq
\int_0^{\omega_m} d\omega \sigma_1(\omega)=\frac{\pi}{2}e^2
(\frac{n_\uparrow}{m^*_\uparrow}+
\frac{n_\downarrow}{m^*_\downarrow})
\eeq
and changes in optical absorption will occur both as a function
of temperature and of magnetic field due to changes in the degree
of spin polarization and corresponding changes in the effective
masses. We assume that the effective mass is proportional to
the inverse bandwidth, so that
\beq
\frac{m}{m^*_\sigma}=1-2j_1I_\sigma-2j_2I_{-\sigma}
\eeq

Figure 8 shows the typical behavior expected in low frequency optical
absorption, for parameters $j_1=0$, $j_2=2$, $n=0.5$, $k=0$
as an example. The optical absorption decreases for $T$
above $T_c$ as the temperature is lowered, and increases again rapidly below
$T_c$. When a magnetic field is applied the optical absorption increases at all
temperatures, with the largest increase at $T_c$. Figure 9(a) shows the optical
weight versus temperature illustrating this behavior. In figure 9(b) optical
weights for majority and minority spin electrons are shown
separately; when a magnetic field is applied, the optical absorption
of majority spins increases and that of minority spins decreases
but less strongly, giving rise to the change seen in Fig. 9(a).

The dc conductivity is given by
\beq
\sigma=\frac{ne^2\tau}{m^*}
\eeq
and the magnetoresistance in our model will be determined by the change
in effective mass, or bandwidth, with spin polarization. Assuming
a constant relaxation time the magnetoresistance is given by
\beqn
\frac{\Delta \rho}{\rho}&\equiv &\frac{\rho(H)-\rho(0)}{\rho(0)}\nonumber \\
&=&
\frac{\sum_\sigma n_\sigma(H)(1-2j_1I_\sigma(H)-2j_2I_{-\sigma}(H))}
{\sum_\sigma n_\sigma(0)(1-2j_1I_\sigma(0)-2j_2I_{-\sigma}(0))} -1
\eeqn
and is shown in Fig. 10 for one case. It is maximum at $T_c$ and remains
large well above $T_c$. Similar results are found for other parameters 
in the model.

Note that under the assumption that there is no significant change in the
relaxation time with spin polarization, there should be a definite relation
between the optical weight, 
\beq
W(T,H)\equiv \frac{2}{\pi e^2}\int_0^{\omega_m} d\omega \sigma_1(\omega)
\eeq
and the magnetoresistance. For example, the quantity
\beq
f(T,H)=\frac{\Delta \rho /\rho(0)}{\Delta W/W(H)}
\eeq
should be a constant independent of temperature and magnetic
field. Deviations from constant behavior would indicate a
dependence of the relaxation time on spin polarization.

Another important experiment that can shed light on the nature
of the ferromagnetic transition is angle-resolved photoemission. For example,
some results on manganites have recently been reported\cite{dess}.
In the models discussed here the quasiparticle dispersion changes
with magnetization, and hence one would expect characteristic
signatures in the photoemission spectrum as function of temperature
or magnetic field. Of particular interest would be to obtain high
resolution angle- $and$ spin-resolved photoemission spectra of ferromagnetic
metals\cite{photo}. Figure 11 shows an example of predicted angle-resolved
photoemission spectrum for our model for one case without
exchange: for majority spins, the Fermi velocity increases as the
system polarizes and the quasiparticle peak disperses faster,
and the opposite occurs for minority spins. For the non-spin-resolved
spectrum (Fig. 11c), two peaks develop as the system polarizes, one of
which disperses slower and the other one faster than in the
unpolarized case. However, for other parameters in the model, in the
presence of exchange, one finds that both majority and minority
spins disperse faster in the ferromagnetic state, reflecting the
gain in kinetic energy upon spin polarization.

\section{Summary and discussion}

We have considered a model for ferromagnetism where the magnetic order
arises from a modification of the width of the bands upon
spin polarization instead of the usual exchange splitting. 
Thus, ferromagnetism arises from a gain in kinetic rather
than potential energy. This represents the opposite extreme of
the conventional understanding of ferromagnetism as arising from a gain in
potential energy (exchange energy) despite an associated cost
in kinetic energy\cite{matt}. More generally, the models considered here
also include an exchange splitting parameter, in the presence of
which ferromagnetism arises from a combination of gains in potential
and kinetic energy.

As we have seen, these models can be derived from a single band
tight binding model that includes off-diagonal matrix elements
of the Coulomb interaction. The terms that give rise to kinetic
energy gain upon spin polarization originate in nearest neighbor 
matrix elements conventionally termed "exchange" and "pair hopping".
Physically, they represent "bond-charge repulsion" \cite{kive,camp}, i.e. the
repulsive  Coulomb energy of electrons at the interstitial region between
neighboring atoms rather than at the atoms themselves. In the present context
the use of the term "exchange" or "Heisenberg exchange" for such matrix
elements is somewhat misleading, because the driving force for ferromagnetism
in fact is direct Coulomb repulsion of electrons in the bonds. The model
discussed in this paper with exchange parameter $k=0$ describes ferromagnetism
without exchange splitting, even though the main interaction is the parameter
$J$ which is conventionally called an exchange integral.

It is generally believed that band degeneracy is essential to
metallic ferromagnetism\cite{matt}. It is certainly true that ferromagnetism
usually arises is systems that have atoms with valence electrons in 
degenerate orbitals. Nevertheless, it is difficult to see how
atomic orbital degeneracy could be essential for example in 
the case of Ni and Ni-Cu alloys, which have a small fraction of 
a d-hole per atom and negligible polar fluctuations. Furthermore,
it should be kept in mind that all energy bands in metals are in
fact non-degenerate, except at sets of states of measure zero in
the Brillouin Zone (points or lines). Hence we believe that an approach
that focuses on a single non-degenerate band, as in this paper,
is sensible, whether that band arises from degenerate or non-degenerate atomic orbitals.
On transforming from Bloch to Wannier states for that band and
computing the matrix elements of the Coulomb interaction with the
Wannier states, the matrix elements discussed in Sect. III (as well as
others) result. The question of what is the magnitude of these matrix 
elements in particular materials is a difficult one. Within the
point of view of this paper (and our previous work\cite{hirs1,hirs6}), band
degeneracy may be important in determining the magnitude of these matrix
elements, but not in determining the structure of the theory.

The model considered here naturally gives rise to partial spin polarization,
even with a constant density of states, unlike the Stoner model. In the
absence of exchange splitting, we found that ferromagnetism does occur,
however full spin polarization can only be achieved in a limited range
of parameter space (far from the half-filled band). As the exchange 
splitting increases from zero, full polarization can be achieved over
an increasing range of band filling. The condition for onset of spin
polarization may become more or less stringent as the band filling
increases depending on the ratio of the same-spin to opposite-spin
band renormalization parameters, $j_1/j_2$.

The magnetization versus temperature showed unconventional behavior in
the absence of exchange, with the curves being substantially less
steep. An interesting feature is that re-entrant ferromagnetism can
occur in certain parameter ranges. The magnetic susceptibility
versus temperature showed Curie-Weiss behavior for a wide range of parameters,
although deviations can also occur.

Re-entrant ferromagnetism, and in particular a situation where the magnetic order
$increases$ as the temperature increases, is somewhat counterintuitive but by
no means unphysical. In fact, experimental observations of such
behavior have been reported in $Y_2Ni_7$\cite{reen1}
 and in $ThFe_3$\cite{reen2}.
Theoretical models that have been found to exhibit such behavior are Ising 
models with random ferromagnetic and antiferromagnetic exchange\cite{reen3},
and even the Stoner model for special forms of the density of
states\cite{reen4}. However in the Stoner model such behavior is always
associated with first order transitions.
In the models discussed here, the re-entrant behavior can
be easily understood. Consider a case where the magnetism has disappeared
at sufficiently low temperatures. As the temperature is raised, the
contribution of entropy to the free energy of the system increases
relative to that of energy. Now in our model (in the absence of exchange), by
spin polarizing one of the bands will narrow and as a consequence the entropy
contribution from electrons in that band will increase, so that this effect
added  to the energy gain from the electrons in the broadened band can lead to
an overall decrease of the free energy upon spin polarization at higher
temperatures. As the temperature increases further there will be a point where
entropy from both bands will favor a vanishing of the magnetization as in the
usual cases.

The most characteristic feature of the models discussed here, 
however, both in the absence and in the presence of exchange splitting, arises
in optical properties. A transfer of optical spectral weight from high
frequencies to low frequencies always occurs upon spin polarization in the
presence of the interactions $j_1$ and $j_2$ that modify the bandwidth. 
Furthermore, above $T_c$ these interactions cause spectral weight to be
transfered from low to high frequencies as the temperature is lowered. These
phenomena have been observed in manganites\cite{okim} and hexaborides\cite{degi}
as a function of temperature. Within the models considered here, the
same effects should be seen as a function of magnetic field. 
In fact, recent
optical experiments on Gd-doped Si in the presence of large magnetic fields
show just such a remarkable behavior\cite{dimi}.A systematic
search for these effects in other ferromagnets may reveal that they are a
universal signature of metallic ferromagnetism, which would lend support
to the models discussed here. The magnitude of the effects however,
just as the magnitude of the magnetoresistance, is likely to vary
widely from material to material. As discussed in Sect. VII, there
should be a definite relation between the magnitude of the magnetoresistance
and the change in optical absorption with magnetic field, if indeed the
dominant effect is the lowering of effective mass with spin polarization,
as suggested by our model, rather than a change in the relaxation time.
Further discussion of these issues and the relation of our model with the
'double exchange model'\cite{doub} is given elsewhere\cite{hirs5}.

Unfortunately, the model discussed here does not in itself contain
the physics of the high energy degrees of freedom from where the
extra optical spectral weight that appears at low frequencies gets
transfered from. It would be of great interest to have a model that
would describe this high energy physics and give rise to the
Hamiltonian considered here as an effective Hamiltonian for its
low energy degrees of freedom. 

Finally, photoemission experiments, especially if angle- and spin-resolved,
should be able to provide essential clues on the validity of the
model discussed here. An increase in the Fermi velocity of electrons of
at least one spin orientation as function of increasing magnetic field
or decreasing temperature (below $T_c$) would be expected to also be a universal
feature of metallic ferromagnets if the models discussed here are applicable.

There has recently been substantial interest in re-analyzing the problem
of metallic ferromagnetism\cite{tasa,voll}, and there seems to be a consensus
that this old problem is still not well understood despite the practical
successes of spin-density-functional theory\cite{moru}. In particular, 
recent work has suggested that peaks in the density of states are the dominant
mechanism giving rise to ferromagnetism\cite{tasa,voll}. It has also been
suggested\cite{voll} that because ferromagnetism is a strong coupling problem
it is not really possible to pinpoint the single ultimate cause of metallic
ferromagnetism, whether peaks in the density of states, band degeneracy, or
particular electron-electron interactions. However, the work discussed here and
previously\cite{hirs1,hirs2,hirs5} suggests otherwise. We expect metallic
ferromagnetism to be always accompanied by the transfer of optical spectral
weight discussed above, as well as by a change in the quasiparticle
dispersion. Establishing experimentally that these phenomena do not occur
would  prove the invalidity of the model discussed here. Concerning band 
degeneracy, if ferromagnetism is found in a system where conduction clearly
occurs through non-degenerate bands, such as metallic hydrogen\cite{hirs4}, it
would establish that band degeneracy is not essential and lend further support
to the model discussed here.

\acknowledgements

I am grateful to D. Basov for sharing results of unpublished experiments.

\begin{figure}
\caption {Density of states for up and down electrons in the
unpolarized (left) and magnetic (right) states (schematic). The dashed 
line indicates the position of the Fermi level. (a) Stoner model:
as the temperature is lowered, the bands for up and down spin electrons
rigidly shift with respect to each other. (b) Model considered
in this paper: as the temperature is lowered, the bands change
their width relative to each other, without relative displacement
of their centers. For band filling above one-half, as shown in the
figure, the broader band corresponds to minority spin carriers;
for band filling below one-half, the situation is reversed.
}
\label{Fig.1}
\end{figure}

\begin{figure}
\caption {Ground state phase diagrams. $a=j_1/j_2$.  $j=j_2$. P, PF and F
denote paramagnetic, partially ferromagnetic and fully
polarized ferromagnetic regions respectively. In the absence
of exchange ($k=0$), full polarization cannot
be achieved for band fillings between $1/4$ and $3/4$; that portion of the phase
boundary is indicated by a dashed line. As $k$ increases the region
where full polarization can be achieved increases and for $k\geq 0.5$
it covers the entire phase diagram except for the point $n=1$.
}
\label{Fig.2}
\end{figure}

\begin{figure}
\caption {Magnetization versus temperature for band filling
$n=0.5$, $k=0$, $j1=0$. The numbers next to the curves give the
values of $j_2=j$. For $j<1.6$ re-entrant ferromagnetism occurs
in this case.
}
\label{Fig. 3}
\end{figure}

\begin{figure}
\caption {Effective mass ratio, or inverse of bandwidth, versus temperature
for the parameters of Fig. 3. (a) Majority spins; (b) minority spins.
The numbers next to the curves give the values of $j_2=j$. 
}
\label{Fig. 4}
\end{figure}

\begin{figure}
\caption {Effective moment (Eq. (40)) versus temperature for the
parameters of Fig. 3.
}
\label{Fig. 5}
\end{figure}

\begin{figure}
\caption {(a) Magnetization versus temperature and (b) effective moment versus
temperature for $n=0.5$, $k=0$, $a=j_1/j_2=0.5$. The values of $j_2=j$ are
given next to the curves. } 
\label{Fig. 6}
\end{figure}

\begin{figure}
\caption {Effect of exchange splitting on the behavior of (a) magnetization
and (b) effective moment versus temperature. For (a), the curves of increasing
steepness correspond to the cases (i) $j/(1-k)=2, k=0$; (ii) $j/(1-k)=2, k=0.25$;
(iii) $j/(1-k)=2, k=0.5$; (iv) $j/(1-k)=4, k=0.5$.
}
\label{Fig. 7}
\end{figure}

\begin{figure}
\caption {Optical conductivity (arbitrary units) versus frequency
for the parameters of Fig. 3. The full lines give the optical 
conductivities at temperatures $T/T_c=0.2$, $1$ and $3$, and the dashed
lines the corresponding ones in the presence of a magnetic field
$h=H/D=0.02$. The effect of the magnetic field is largest at
$T_c$, and becomes very small for $T/T_c=3$.
}
\label{Fig. 8}
\end{figure}

\begin{figure}
\caption {Optical weight $n/m^*$ given by the low-frequency integral
of the conductivity versus temperature in the presence of a magnetic
field $h=H/D$ (numbers next to the lines). (a) Total, (b) for spin up
(full lines ) and spin down (dashed lines) respectively. The parameters are those
of Fig. 3.
}
\label{Fig. 9}
\end{figure}

\begin{figure}
\caption {Magnetoresistance versus temperature for the parameters of Fig. 3.
}
\label{Fig. 10}
\end{figure}

\begin{figure}
\caption {Angle-resolved photoemission spectra for the parameters of
Fig. 3, giving rise to $T_c/D=0.081$. We assume a bandwidth
$D=0.3 eV$, so $T_c=280K$. Full lines correspond to temperature
$T/T_c=1.2$, dashed lines to $T/T_c=0.8$. The numbers next to
the full lines give the values of $\epsilon_k-\mu$ (in meV).
(a) Majority spins; below $T_c$, the Fermi velocity increases and the
peaks disperse faster. (b) Minority spins: the Fermi velocity decreses
upon ordering and the peaks disperse slower. (c) Non-spin-resolved:
two peaks appear below $T_c$, one disperses faster and one slower than
the peaks above $T_c$. Similar results would be found at fixed
temperature under application of a magnetic field. In particular,
application of a magnetic field $h=H/D=0.03$ at $T/T_c=1.2$ gives
results almost identical to those shown in the figure for $T/T_c=0.8$.
}
\label{Fig. 11}
\end{figure}

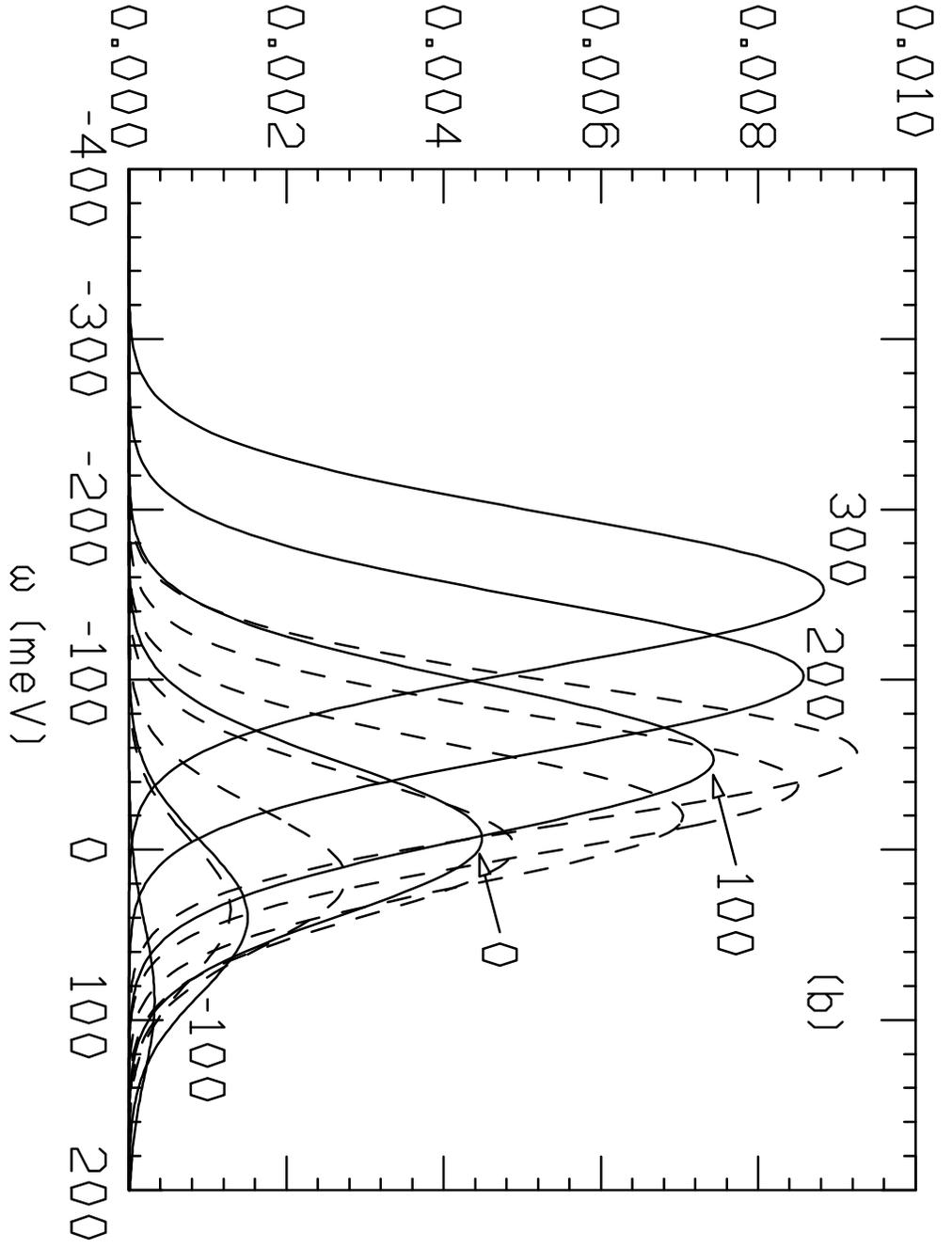
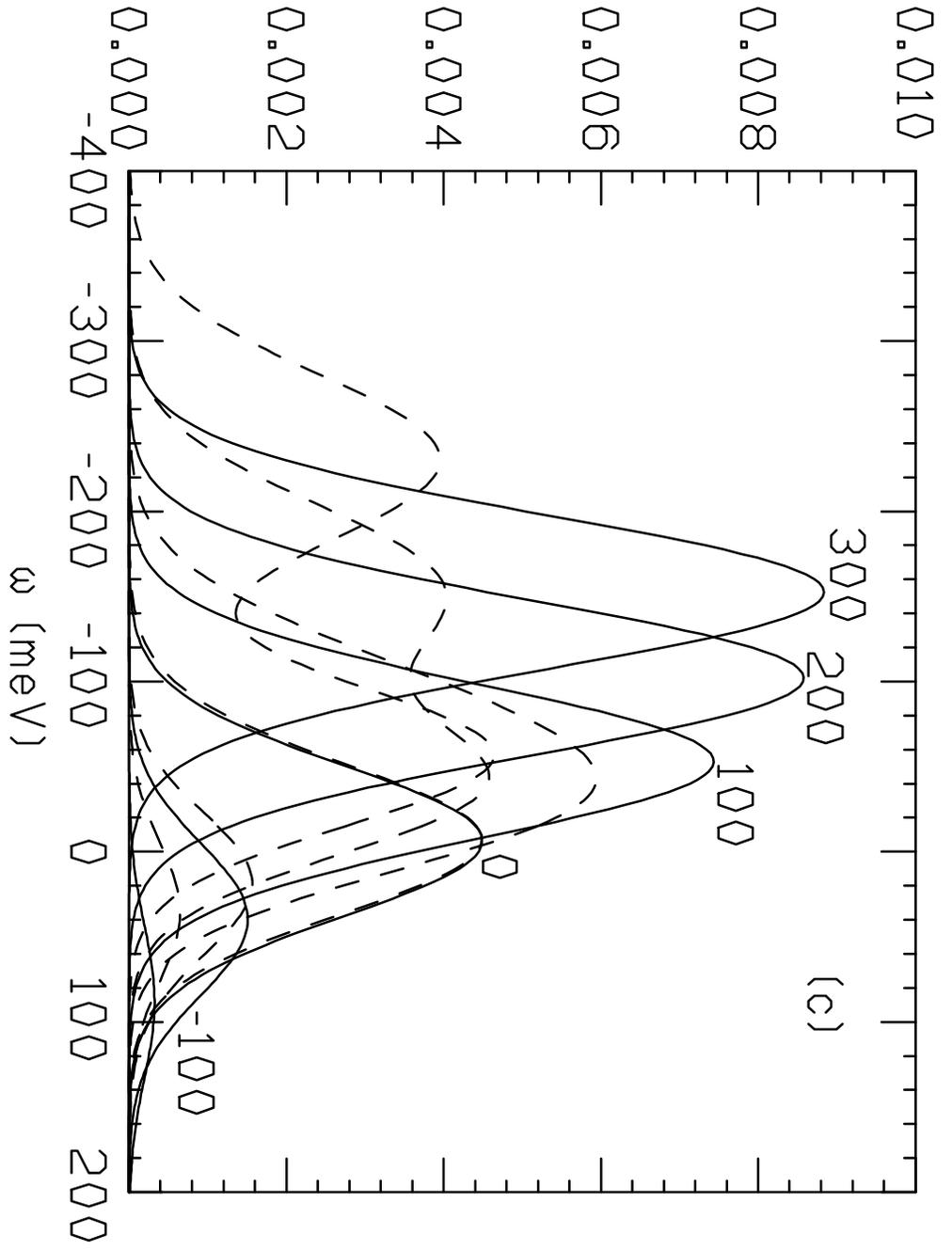
\end{document}